\documentclass[11pt]{article}

\usepackage{xcolor,fullpage}
\usepackage{pslatex}
\usepackage{graphicx}
\usepackage{fancyhdr}
\usepackage{amsmath}
\usepackage{amssymb}
\usepackage{textcomp}
\usepackage[square,sort,comma,numbers]{natbib}
\usepackage{subcaption}	
\usepackage{hyperref}
\usepackage{setspace}
\usepackage{cancel}
\usepackage{mathtools}
\usepackage[font=footnotesize]{caption}
\usepackage{rotating}
\usepackage{adjustbox}
\usepackage{multirow}
\usepackage{tabularx}
\usepackage{lipsum}
\usepackage{booktabs}
\usepackage{subfiles}
\usepackage{titling}
\usepackage{placeins}
\usepackage{authblk}
\usepackage{comment}
\usepackage{wrapfig}

\newcolumntype{L}[1]{>{\raggedright\let\newline\\\arraybackslash\hspace{0pt}}m{#1}}
\newcolumntype{C}[1]{>{\centering\let\newline\\\arraybackslash\hspace{0pt}}m{#1}}
\newcolumntype{R}[1]{>{\raggedleft\let\newline\\\arraybackslash\hspace{0pt}}m{#1}}

\title{Extending the aircraft flight envelope by mitigating transonic airfoil buffet}

\author[1,2]{Esther Lagemann$^{\dagger}$} 
\author[1]{Steven L. Brunton}
\author[2]{Wolfgang Schröder}
\author[1,2]{Christian Lagemann$^{\dagger}$} 
\affil[1]{\small AI Institute in Dynamic Systems, Department of Mechanical Engineering, University of Washington, Seattle, WA 98195, United States}
\affil[2]{\small Institute of Aerodynamics, RWTH Aachen University, W\"ullnerstra\ss e 5a, 52062 Aachen, Germany}
\affil[$^{\dagger}$]{\small Authors contributed equally.}

\begin{document}
\date{}
\maketitle

\begin{abstract} 
In the age of globalization, commercial aviation plays a central role in maintaining our international connectivity by providing fast air transport services for passengers and freight. However, the upper limit of the aircraft flight envelope, i.e., its operational limit in the high-speed regime, is usually fixed by the occurrence of an aerodynamic phenomenon called transonic airfoil buffet. It refers to shock wave oscillations occurring on the aircraft wings, which induce unsteady aerodynamic loads acting on the wing structure. Since these loads can cause severe structural damage endangering flight safety, the aviation industry is highly interested in suppressing transonic airfoil buffet to extend the flight envelope to higher aircraft speeds. In this contribution, we demonstrate experimentally that the application of porous trailing edges substantially attenuates the buffet phenomenon. Since porous trailing edges have the additional benefit of reducing acoustic aircraft emissions, our findings could pave the way for faster air transport with reduced noise emissions. \footnote{A video research abstract can be found at: \url{https://www.youtube.com/watch?v=ddomrz1cz_U&t=1s}}
\end{abstract}

\section{Introduction}
Each aircraft has an individual flight envelope that ensures safe operations. It is typically determined during the design phase, as it defines operational limits with respect to aircraft speed, load factor, i.e., the ratio of lift to weight, and atmospheric conditions. Operations outside this envelope can result in severe structural damage, endangering flight safety and must be avoided at any time~\cite{jacquin2009experimental,giannelis2017review,xu2023}. \\
For modern lightweight aircraft, the high-speed limit of the flight envelope is usually defined by the onset of an aerodynamic phenomenon called transonic airfoil buffet~\cite{crouch2009origin,korthauer2023,iwatani2023}. Even though the aircraft is operated at subsonic speeds, i.e., the Mach number $Ma$ is smaller than 1, the airfoil's shape causes an acceleration along the front part of its upper surface that creates a local supersonic region ($Ma > 1$). This region is terminated by a shock wave, which abruptly decelerates the flow to subsonic speed by an air compression.
For particular flow conditions, which are defined by certain combinations of Mach number, Reynolds number, and angle of attack, the shock wave becomes unsteady and oscillates in the streamwise direction~\cite{schauerte2023,kojima2020,paladini2019analysis}. This self-sustained shock wave oscillation is called transonic airfoil buffet. Although extensive experimental and numerical investigations have been conducted to understand the underlying mechanism(s) of this aerodynamic phenomenon~\cite{deck2005numerical,jacquin2009experimental,hartmann2013interaction,feldhusen2018investigation,paladini2019various,sartor2015stability,accorinti2022experimental,sansica2022,poplingher2019modal,moise2023transonic}, its root causes and self-sustaining mechanisms are still controversially discussed~\cite{giannelis2017review}. Amongst the variety of hypotheses, one theory suggests that vortical structures propagate from the shock wave downstream to the trailing edge, where they generate acoustic waves that travel upstream and interact with the shock wave~\cite{lee1990}. Various experimental measurements reported good agreement with this theory, e.g., \cite{hartmann2013interaction,feldhusen2018investigation,feldhusen2021analysis,dAguanno2021experimental}. Another hypothesis is based on the close relation between the shock wave oscillation and the time-dependent variation of the shock-induced separation, which essentially modifies the pressure ratio across the shock wave in a time-dependent manner~\cite{iovnovich2012reynolds,fukushima2018wall,iwatani2023}.
However, despite discrepancies in the hypotheses about the underlying physical phenomena, it is unambiguous that transonic airfoil buffet constitutes a severe threat to aircraft operations. The self-sustained shock wave oscillations yield unsteady aerodynamic loads acting on the wing structure~\cite{raveh2011frequency,raveh2014aeroelastic,korthauer2023experimental} and potentially result in structural failure. Consequently, the respective flow conditions are excluded from the flight envelope. However, to enable faster and safe air transport, attenuating transonic airfoil buffet is of substantial interest in aerospace engineering. \\[0.2cm]
Therefore, it is not surprising that different strategies have been developed to optimize the high-speed aerodynamic performance by influencing the shock wave. Broadly, they can be divided into passive and active control methods. Passive approaches have the potential benefit of simplicity, robustness, low weight, and ease of retrofit~\cite{eastwood2012}. While active methods are usually more complex and require a higher level of maintenance, their real-time feedback control is expected to provide better off-design performance compared to passive technologies.\\[0.2cm]
A straightforward passive approach is to increase the structural damping. Since this comes at the cost of weight increase and thus, higher fuel consumption, it is certainly not the most efficient solution. Another passive approach is the installation of small bumps on the upper wing surface. These shock control bumps were shown to weaken shock waves, resulting in wave drag reduction~\cite{ashill1992}, and to improve the buffet margin~\cite{eastwood2012,mayer2019}. However, they are highly sensitive to the flow conditions including shock strength, shock position, and post-shock pressure gradient~\cite{bruce2015}. For instance, an incorrect placement relative to the shock wave position results in the undesired effect of even stronger shock waves~\cite{ogawa2006}, which degrades the aerodynamic performance in off-design operations.\\[0.2cm]
Another family of control devices, which comprises passive and active technologies, relies on the generation of vortices that energize the boundary layer via momentum transfer. The resulting stabilization of the boundary layer is directly linked to a shock wave stabilization, which was observed for mechanical, plasma, and fluidic vortex generators with pulsed and continuous blowing~\cite{huang2012,molton2013,sidorenko2019}. Although active devices possess the advantage of adjustability with respect to changing flow conditions via feedback control, the fixed location along the airfoil limits a flexibility with respect to off-design conditions for all vortex generators. Moreover, they increase drag~\cite{huang2012,sidorenko2019}, which degrades the aerodynamic performance. 

\begin{figure}[t]
\includegraphics[width=1.\textwidth]{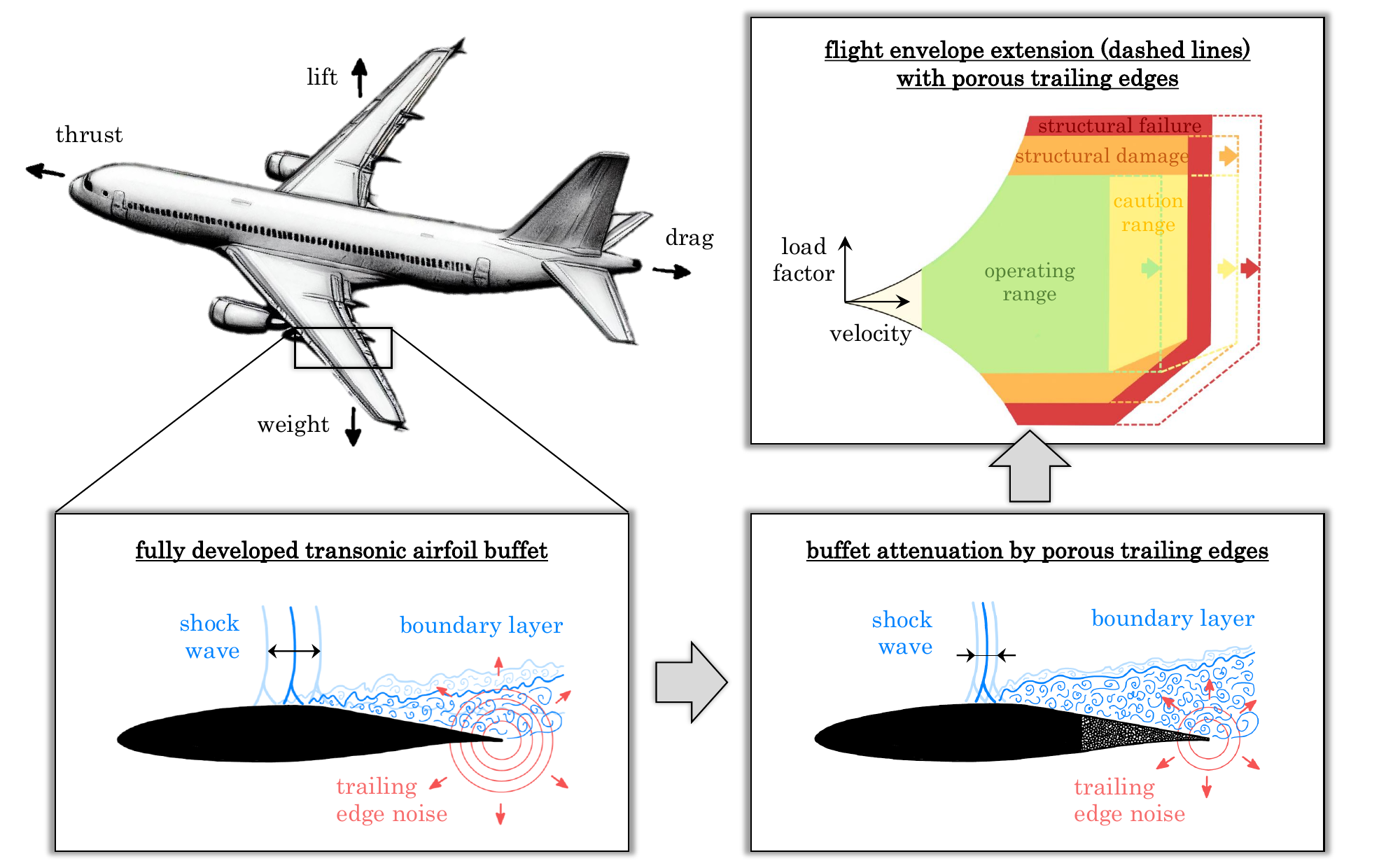}
  \caption{\textbf{Buffet attenuation and flight envelope extension by porous trailing edges.} Fully developed transonic airfoil buffet describes self-sustained shock wave oscillations along the suction side, i.e., the upper side, of the airfoil. Since this shock movement imposes severe structural loads that potentially result in structural failure of the aircraft wing, the typical aircraft flight envelope is limited to aircraft speeds below the buffet onset boundary. We demonstrate experimentally that the installation of porous trailing edges attenuates the buffet phenomenon substantially. This allows an extension of the flight envelope to higher velocities without hazarding the aircraft's structural integrity. Such an extension is marked in the flight envelope diagram by dashed lines and arrows. Moreover, porous trailing edges have been shown to reduce the trailing edge noise~\cite{koh2018,ali2018,rossignol2020,zhou2018discrete}. Thus, our approach can lead to faster and safer aircraft operation with reduced noise emission.}
  \label{fig:motivation}
\end{figure}

To decouple the efficiency of control strategies from sensitive flow conditions, such as the shock wave position, research has focused on trailing edge devices. A prominent example are trailing edge deflectors and flaps, which are incorporated into a closed loop control system that moves the devices in response to the shock wave. By counteracting the shock movement with trailing edge pressure changes induced by the flap movement, these devices can successfully control transonic buffet~\cite{caruana2005,gao2016,ren2020,zhang2022,gao2017active}. Their passive counterpart, i.e., a flap with a fixed deflection angle, is at least able to reduce the amplitude of the shock wave oscillations~\cite{lee1992,aguanno2023}. Such trailing-edge flaps are usually used as high-lift devices to improve the aerodynamic performance during take-off and landing. 
Consequently, a trade-off must be found with respect to a trailing-edge flap design that meets both needs - high-lift performance and buffet attenuation. Moreover, the deployment of high-lift components increases the far-field noise level, heavily contributing to the overall aircraft noise during landing~\cite{guo2003}. However, due to the rising evidence of mental and physical health implications associated with aircraft noise \cite{munzel2021,koczorowski2022,nguyen2023}, there is currently a strong demand for lowering acoustic emissions. In this respect, many academic studies have focused on trailing edge modifications since even for the clean wing, i.e., high-lift devices retracted, trailing edges constitute a dominant noise source term ~\cite{guo2003}. It has been shown in numerous studies ~\cite{koh2018,ali2018,rossignol2020,zhou2018discrete} that porous trailing edges can efficiently reduce trailing edge noise.\\[0.2cm]
In this contribution, we will demonstrate experimentally that porous trailing edges additionally constitute a promising technology for transonic airfoil buffet attenuation (see figure~\ref{fig:motivation}). This two-fold benefit is in great contrast to the other buffet control strategies discussed above. These devices typically increase the noise emissions because the additional components disturb the flow field~\cite{satcunanathan2022impact,zamponi2023effect}. Moreover, porous trailing edges have the advantages of low cost, low operational complexity, and robustness. Since they do not directly target the shock wave itself, the sensitivity to the flow conditions and thus, an adverse off-design performance, is minimized. Moreover, these devices are easily manufactured using state-of-the-art 3D selective laser melting techniques, which allows an immense flexibility regarding design aspects such as the airfoil shape or material. Furthermore, we made two conscious design choices to minimize aerodynamic performance losses. First, lift degradation is counteracted by incorporating an impermeable plate inside the porous material. Without this blockage, a mass flux from the pressure to the suction side would yield a pressure compensation that has been shown to lower the aerodynamic lift~\cite{geyer2010,geyer2014,ananthan2020}. Second, a perforated surface layer is added on top of the porous surfaces. This reduces the wall roughness and minimizes turbulence production, which was reported to increase the aerodynamic drag of porous surfaces~\cite{jimenez2001,koh2018,teruna2021}.\\[0.2cm]
Overall, the findings presented in this contribution clearly demonstrate the ability of porous trailing edges to mitigate transonic airfoil buffet. Due to their benefits compared to other buffet control technologies, they possess great potential in advancing civil aviation by extending the limits of the flight envelope in the high-speed regime.

\section{Results}
As a highly customizable and rapidly producible airfoil device, porous trailing edges present a quick, low-cost, and robust yet effective buffet control technique with significant practical advance for the entire aviation sector. To demonstrate experimentally the beneficial properties of porous trailing edges, comprehensive wind tunnel tests have been performed. We start by briefly outlining the experimental setup and the measurement facility. This is followed by a presentation of the flow field characteristics of the unmodified reference case, i.e., an ordinary solid trailing edge, before the buffet attenuation capabilities of our porous trailing edge designs are reported. Subsequently, we discuss the aerodynamic performance of the investigated trailing edge designs. Finally, we elaborate on the physical mechanism by which the investigated porous trailing edges damp the self-sustained shock wave oscillations.

\subsection{Experimental setup}\label{sec:setup}
The experimental investigations are conducted in the \textit{Trisonic Wind Tunnel} facility of the Institute of Aerodynamics at RWTH Aachen University. To investigate the desired flow conditions exhibiting distinct shock wave oscillations on the suction side, i.e., the upper side, of the airfoil, a Mach number of $Ma = 0.76$, an angle of attack of $\alpha = 3.5^\circ$, and a chord-length based Reynolds number of $Re_c = 2.1\cdot 10^6$ are chosen. To put it in perspective, this Reynolds number is at the lower bound of the spectrum for civil aircraft ~\cite{lissaman1983,shirbhate2019}.

\begin{figure}[t]
\includegraphics[width=1\textwidth]{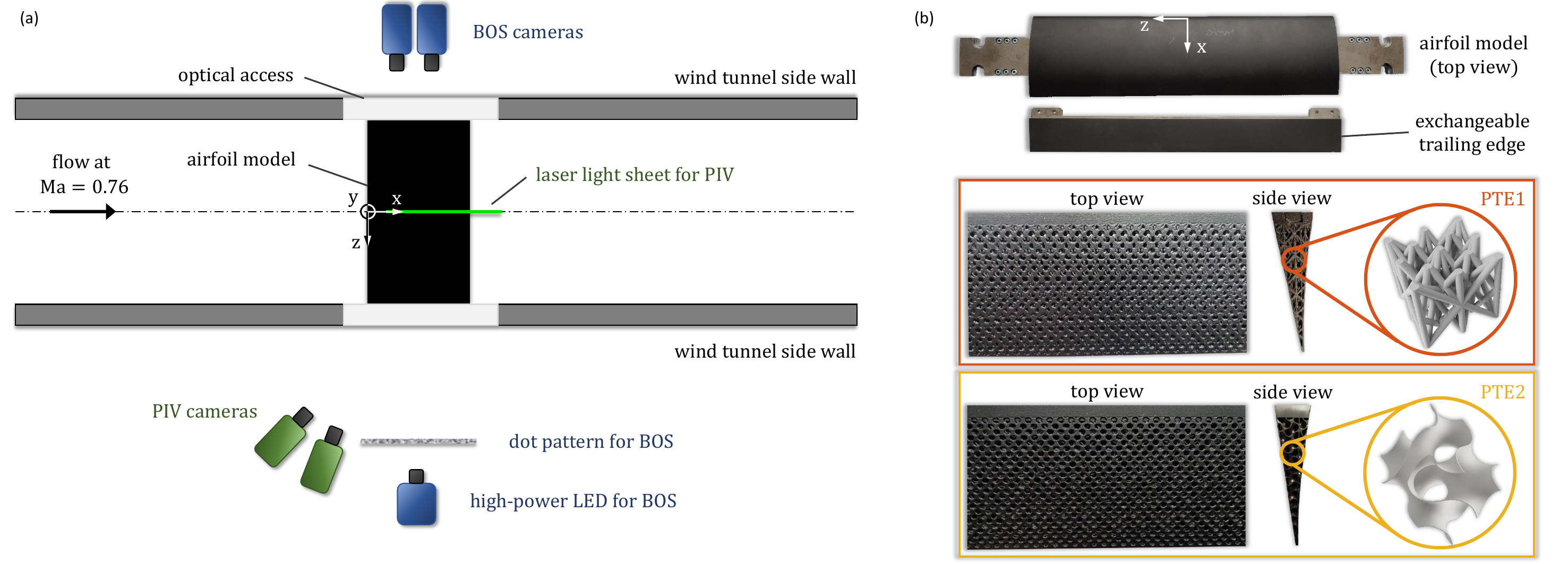}
  \caption{\textbf{Experimental setup.} (a) The experiments are conducted in the transonic measurement section equipped with a rigid airfoil model. The flow field above the airfoil is visualized by two synchronized measurement systems. The PIV setup captures the velocity field and the BOS setup measures density gradients, the latter allowing a precise localization of the oscillating shock wave. (b) Pictures of the airfoil model and the exchangeable trailing edges. Two types of porous trailing edges are tested: PTE1 is based on a three-dimensional lattice structure and PTE2 comprises stacked gyroid cubes.}
  \label{fig:setup}
\end{figure}

The rigid airfoil model's shape is based on the supercritical DRA $2303$ profile, which was previously investigated in, e.g., \cite{hartmann2012, hartmann2013interaction, hartmann2013coupled, feldhusen2018investigation, feldhusen2021analysis}. The model possesses an exchangeable trailing edge section to facilitate the study of different configurations, i.e., an ordinary solid and two porous trailing edges. The original solid trailing edge configuration is denoted as reference case (REF) in the following, whereas the porous trailing edges are abbreviated by PTE1 and PTE2. The PTEs consist of different deterministic structures as depicted in figure~\ref{fig:setup}. To be more precise, PTE1 is based on a three-dimensional lattice structure and PTE2 comprises stacked gyroid cubes. To prevent flow perturbations generated at sparsely connected struts, a perforated surface layer is added to both PTEs, which lines up precisely with the surface contour of the leading edge part. Moreover, a solid plate along the centerline of the profile prevents artificial mass flux and pressure compensation between suction and pressure side of the airfoil to avoid aerodynamic performance degradation. Both porous trailing edges consist of titanium and are 3D-printed as a single part leveraging state-of-the-art selective laser melting processes. 

To accurately resolve relevant quantities of the transonic airfoil flow in space and time, i.e., density gradient and velocity fields, a sophisticated multi-camera measurement setup is realized comprising simultaneous and synchronized planar high-speed Particle-Image Velocimetry (PIV) and Background-Oriented Schlieren (BOS) measurements (see figure~\ref{fig:setup}). The PIV data provide streamwise and streamwise orthogonal velocity information while the BOS data contain density variations. Since a shock wave constitutes a discontinuous density jump in the flow field, its instantaneous position can be extracted precisely from the BOS data. The PIV data are used to understand how the PTEs influence the flow field in the trailing edge region to attenuate the buffet phenomenon. More details about the experimental setup and data evaluation are provided in the \textit{'Methods'} section.\\
Throughout this paper, the following notation regarding the coordinate system will be used: The $x$-coordinate and the streamwise velocity component $u$ point in the horizontal direction and the $y$-coordinate and the normal velocity component $v$ are orthogonal to the $x$-coordinate as indicated in figure~\ref{fig:setup}. 

\subsection{Buffet characteristics of the reference case} \label{sec:ref}
\begin{figure}[t]
\includegraphics[width=1.\textwidth]{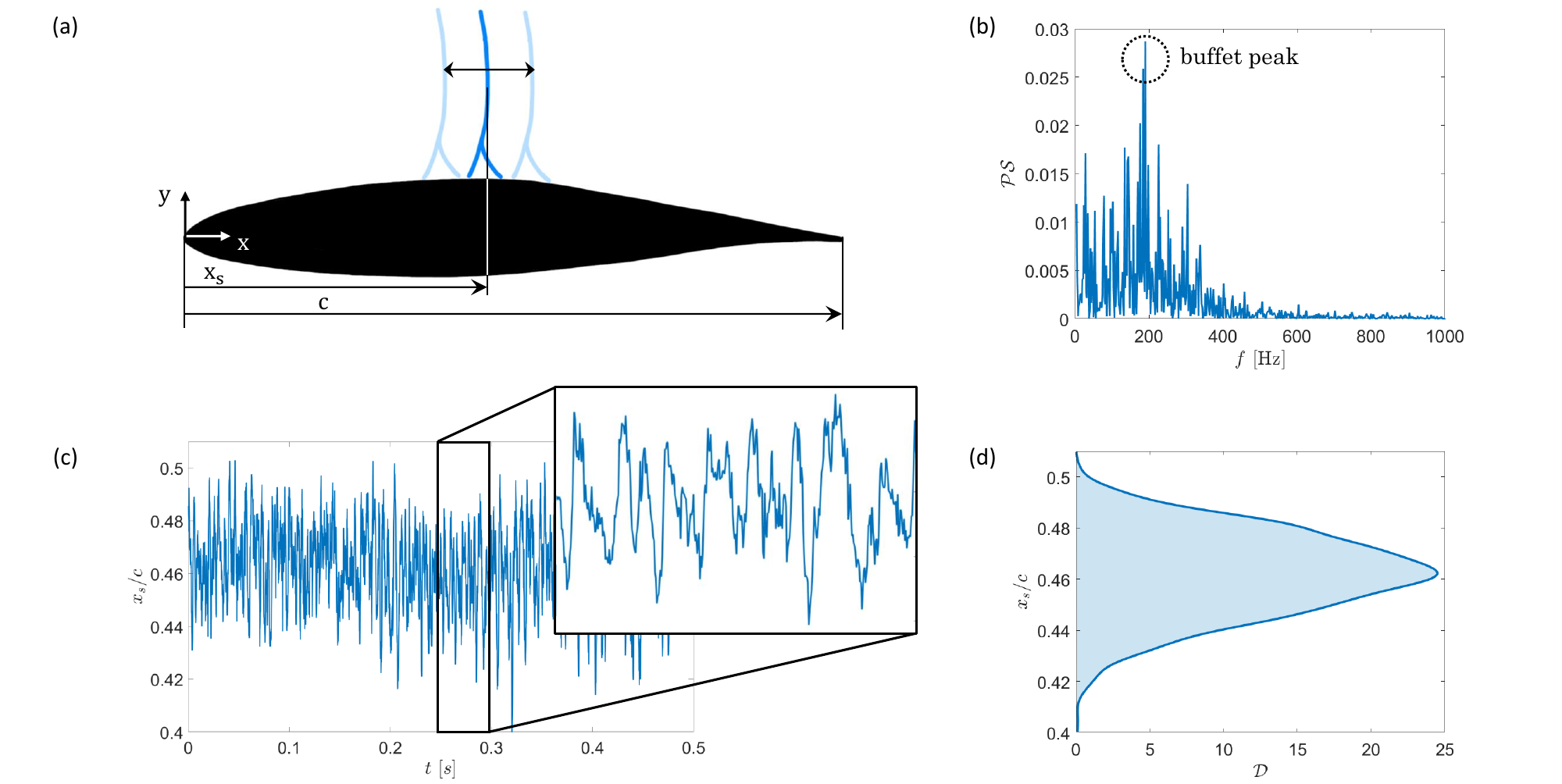}
  \caption{\textbf{Fully developed transonic airfoil buffet of the reference (REF) configuration.} A sketch of the shock wave oscillation along the airfoil is given in (a), while (b-d) show measurement data related to the shock wave movement. The instantaneous shock wave position normalized by the chord length $x_s/c$ is given in (c) and its PDF in (d). The power spectrum of the shock wave position obtained by applying a Fourier transform is provided in (b). It reveals that the shock wave oscillates at a dominant buffet frequency of $f \approx 180$~Hz.}
  \label{fig:ref}
\end{figure}

Fully developed buffet characteristics describe a self-sustained shock wave oscillation along the suction side of the airfoil as sketched in figure~\ref{fig:ref}~(a). The time-dependent shock wave position $x_s$ normalized by the chord length $c$ is shown in figure~\ref{fig:ref}~(c). The typical range of motion is about $6$~\% of the chord and, as depicted by the enlargement, it possesses periodic tendencies. Therefore, the buffet frequency peak, i.e., the prevailing frequency of the shock wave oscillation, can be extracted by applying a discrete Fourier transform to the time-dependent shock wave position. The corresponding normalized power spectrum is shown in figure~\ref{fig:ref}~(b) indicating a buffet frequency of $f \approx 180$~Hz in line with previous investigations~\cite{hartmann2012, feldhusen2018investigation, feldhusen2021analysis}. Additionally, figure~\ref{fig:ref}~(d) depicts the probability density function (PDF) of the shock wave position, which comprises the same information as figure~\ref{fig:ref}~(c) but in a time-independent representation. Therefore, it provides easy access to the median shock position as well as its standard deviation. For this reference case, severe shock wave oscillations are observed with a standard deviation of $std(x_s/c) = 0.0159$.\\

\subsection{Buffet attenuation effect of porous trailing edges} \label{sec:pte}
\begin{figure}[t]
\includegraphics[width=1.\textwidth]{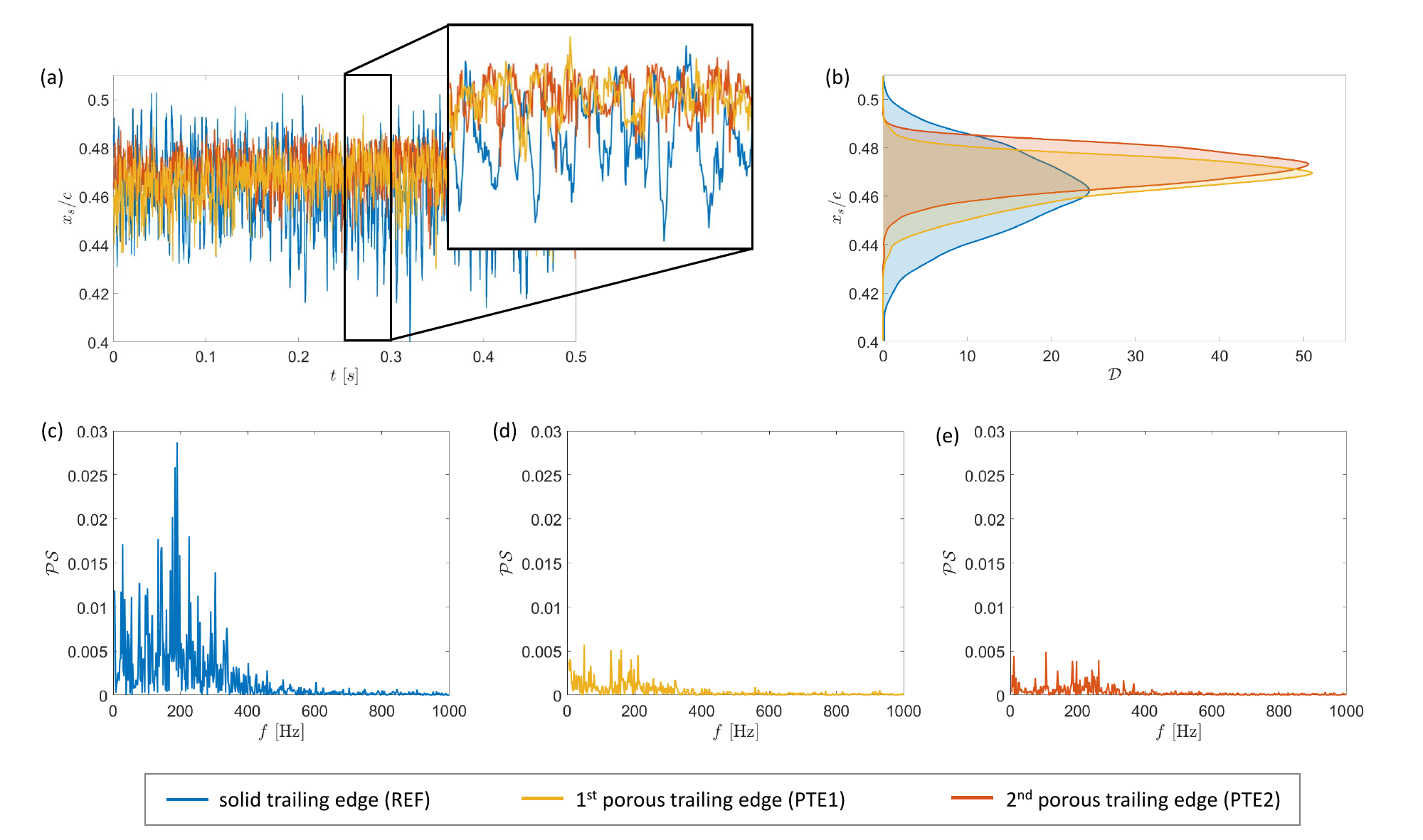}
  \caption{\textbf{Damping shock wave oscillations in fully developed buffet conditions using porous trailing edges.} The instantaneous shock wave position normalized by the chord length $x_s/c$ is given in (a) for both PTEs and the reference case (REF), while the corresponding PDFs are shown in (b). Both representations convincingly demonstrate that both PTE designs successfully attenuate the shock wave movement. The power spectra of the shock wave positions obtained by applying a Fourier transform are provided in (c-e). They clearly reveal that both PTEs (d,e) substantially reduce the energy contained in the shock wave movement, i.e., they damp shock oscillations and eliminate the frequency peak usually associated with transonic airfoil buffet.}
  \label{fig:ptes_shock}
\end{figure}

Next, the influence of two porous trailing edge designs consisting of different deterministic structures on the buffet characteristics is studied. Figure~\ref{fig:ptes_shock} shows the shock properties previously discussed for the reference configuration for all three test cases, i.e., REF, PTE1, and PTE2. The time-dependent shock wave position (figure~\ref{fig:ptes_shock}~(a)) clearly depicts the reduced shock wave oscillations induced by both PTEs. The time-independent representation provided by the corresponding PDFs in figure~\ref{fig:ptes_shock}~(b) convincingly shows narrower distributions with a stronger peak probability for both PTEs, which demonstrates the reduced shock wave oscillations. Moreover, the diagram reveals that the median shock position is moved slightly downstream, i.e., to higher $x_S/c$ values, when the PTEs are deployed. A quantitative comparison of the average shock position $\bar{x}_s/c$ and the corresponding standard deviation $std(x_s/c)$ is provided in table~\ref{tab:properties}. \\
In figure~\ref{fig:ptes_shock}~(c-e), the normalized power spectra of the time-dependent shock position of all three configurations are shown. They clearly demonstrate that both PTEs substantially reduce the energy contained in the shock wave movement, i.e., they damp shock oscillations and eliminate the frequency peak usually associated with transonic airfoil buffet. 

To understand how the PTEs interfere with the flow to attenuate the buffet phenomenon, we investigate the characteristics of the separated boundary layer downstream of the shock wave. Due to the separation location, the PTEs can directly interact with the rear part of this boundary layer flow but not with the shock wave. As recently demonstrated by high-fidelity numerical simulations~\cite{iwatani2023}, the shock movement is directly coupled with the "breathing" of the boundary layer, i.e., a time-dependent thickening and thinning. Although these observations are made using a different airfoil geometry, we will show that the hypothesized mechanism is supported by the flow behavior of the DRA $2303$ airfoil used in our experiments. 

\begin{table}[b]
\footnotesize
\centering
\caption{Average ($\bar{(\cdot)}$) and standard deviation ($std(\cdot)$) of the shock wave position $x_s$ and of the boundary layer thickness $\delta$ normalized by the chord length $c$. Values in parentheses depict the percentage difference to the values of the reference case.}
\label{tab:properties}
\begin{tabular}{c||c|c||c|c}
 & $\bar{x}_s/c$ & $std(x_s/c)$ & $\bar{\delta}/c$ & $std(\delta/c)$ \\
 \hline
REF & $0.4620$ & $0.0159$ & $0.0970$ & $0.0311$ \\
PTE1 & $0.4717 \quad (2.1~\% \uparrow)$ & $0.0075 \quad (52.8~\% \downarrow)$ & $0.1069 \quad (10.2~\% \uparrow)$ & $0.0238 \quad (23.5~\% \downarrow)$\\
PTE2 & $0.4665 \quad (1.0~\% \uparrow)$ & $0.0088 \quad (44.7~\% \downarrow)$ & $0.0988 \quad (1.9~\% \uparrow)$ & $0.0208 \quad (33.1~\% \downarrow)$\\
\end{tabular}\\
\end{table}

\begin{figure}[t]
\includegraphics[width=1.\textwidth]{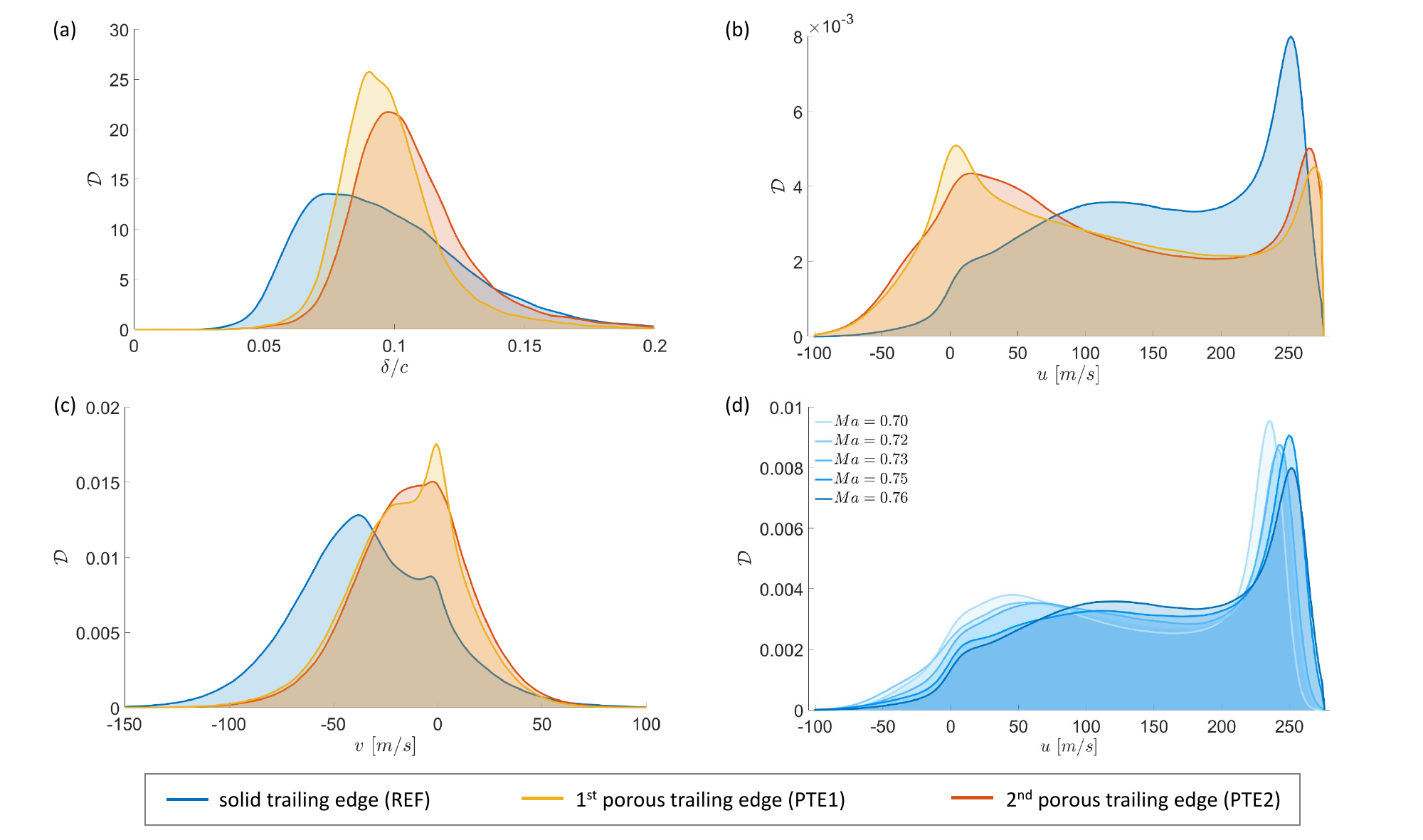}
  \caption{\textbf{Modification of the boundary layer characteristics in fully developed buffet conditions using porous trailing edges.} The PDFs of the boundary layer thickness normalized by the chord length $\delta/c$ of both PTEs and the reference case (REF) are depicted in (a) and clearly reveal the reduced boundary layer breathing in the presence of PTEs. PDFs of the streamwise (b) and the vertical (c) velocity components within the boundary layer show the substantial modifications induced by the PTEs. A comparison to the streamwise velocity PDFs of the reference configuration at different Mach numbers (d) reveals that the PTEs induce a streamwise velocity distribution similar to pre-buffet conditions ($Ma \le 0.72$).}
  \label{fig:ptes_bl}
\end{figure}

Figure~\ref{fig:ptes_bl}~(a) shows a PDF of the boundary layer thickness normalized by the chord length $\delta/c$. The boundary layer thickness is defined as the vertical distance to the airfoil surface at which the streamwise velocity reaches $99~\%$ of the freestream velocity. When PTEs are deployed, the median boundary layer thickness is increased, which is revealed by a shift of the peak probability to larger values of $\delta/c$. Moreover, the time-dependent variation of the boundary layer thickness is reduced, which can be derived from the increased peak strength and the decreased standard deviation in the presence of the PTEs. Table~\ref{tab:properties} provides the respective quantitative values for these observations.  \\
Figure~\ref{fig:ptes_bl} additionally depicts PDFs of the streamwise (b) and the vertical (c) velocity components within the boundary layer for all three test cases. It is obvious that both PTEs substantially change the velocity distribution in the boundary layer. The streamwise velocities (b) are shifted to smaller values, i.e., larger recirculation regions with reversed flow occur, although the velocity at the edge of the boundary layer reaches the same magnitude as in the reference configuration. The PDFs of the vertical velocity component (c) indicate a shift to larger velocity values in the presence of PTEs. To be precise, they are more symmetric around zero. \\
We noticed that the streamwise velocity distributions in the presence of PTEs resemble the observations in pre-buffet conditions of the unmodified reference case. Therefore, figure~\ref{fig:ptes_bl}~(d) shows the respective PDFs related to the reference configuration for several Mach numbers. With the onset of buffet around $Ma = 0.73$, a re-distribution from low to high streamwise velocities is observed. The PDF's peak at velocities close to zero vanishes and only the high-velocity peak at the boundary layer edge remains. The pre-buffet distributions ($Ma \le 0.72$) with a double-sided peak are qualitative comparable to the PTE related distributions at fully developed buffet indicating the substantial importance of the streamwise velocity distribution inside the boundary layer for the buffet mechanism.

\subsection{Aerodynamic performance optimization}
\begin{figure}[t]
\includegraphics[width=1\textwidth]{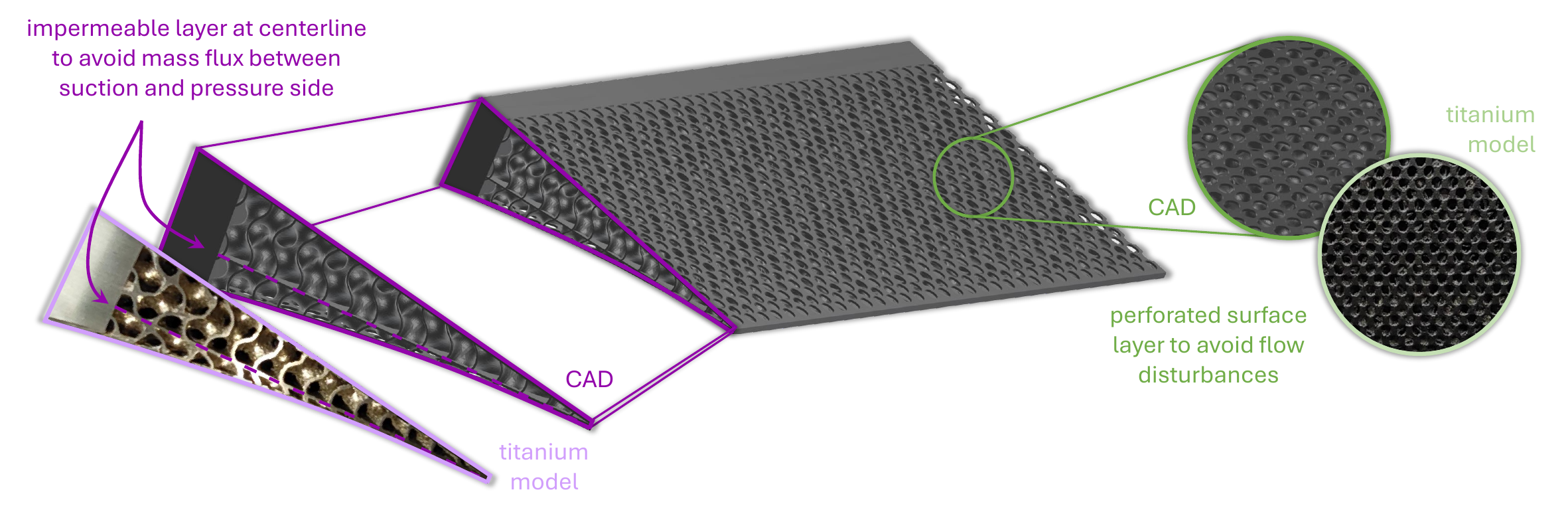}
  \caption{\textbf{Model design choices that optimize the aerodynamic performance of the porous trailing edge configurations.} When mitigating transonic airfoil buffet to extend the flight envelope in the high-speed regime, it is important to consider the effects of the respective airfoil modifications on the aerodynamic performance. To prevent a lift loss, an impermeable solid layer is added at the centerline. This blockage inhibits a mass flux between pressure and suction side, which would yield a pressure compensation degrading the aircraft's lift force. Moreover, a perforated surface layer is added on both sides of the porous trailing edges. Since the porous material constitutes very rough surfaces, they typically disturb the near-wall flow field massively resulting in increased turbulence intensity. This additional viscous drag would increase the overall aircraft's drag requiring a higher fuel consumption. Thus, both design choices aim at counteracting the aerodynamic performance penalties in the presence of porous trailing edges. The figure shows two representations of the PTE2 configuration, i.e., the Computer-Aided Design (CAD) model and photographs of the titanium model.}
  \label{fig:aerodynamics}
\end{figure}

The introduction of a novel technique for transonic airfoil buffet mitigation requires the investigation of its influence on the overall aerodynamic performance. In this respect, we made two conscious design choices, which are highlighted in figure~\ref{fig:aerodynamics}, to minimize adverse effects on the aircraft's aerodynamics. First, we incorporate an impermeable titanium layer at the centerline, which prohibits any mass flux between pressure and suction side of the airfoil. Such a mass flux would result in a pressure compensation with adverse effects on the lift force. This performance degradation is extensively reported in studies investigating porous materials for noise reduction~\cite{geyer2010,geyer2014,ananthan2020}. Inserting an impermeable separation layer inside the porous trailing edges avoids such a lift degradation. Second, we add perforated surface layers on both sides of the porous trailing edges, which are perfectly joined with the solid surfaces of the front airfoil part. The purpose of these perforated surfaces is to provide a smooth surface for the flow field. Since previous investigations of porous trailing edges for noise reduction have noted a significant increase in turbulence intensity due to the roughness of the porous surfaces~\cite{jimenez2001,koh2018,teruna2021}, we minimize this undesired effect by reducing the wall roughness substantially. 

To ultimately prove the effectiveness of these design aspects, complementary measurements of lift and drag forces based on a high-precision force balance are required. Unfortunately, these measurements were not feasible since the wind tunnel facility does not offer such a measurement system for transonic flow conditions. However, we found convincing indications by inferring the aerodynamic performance from other flow quantities suggesting that our carefully chosen designs behave as intended and presumably counteract aerodynamic performance losses. \\
The most important contributor to an increased drag force that could potentially be modified by porous trailing edges is the viscous drag. It is primarily generated by the boundary layer downstream of the shock wave. As figure~\ref{fig:ptes_bl}~(a) and table~\ref{tab:properties} demonstrate, the averaged boundary layer thickness is increased in the presence of porous trailing edges, which may induce a higher viscous drag as shown in \cite{geyer2010,geyer2010porous,satcunanathan2022impact,teruna2020,teruna2021,teruna2021numerical}. These studies report that the increased drag is typically accompanied by a higher turbulence stress causing the aerodynamic performance deficit. On the contrary, our sophisticated design influences the Reynolds shear stress and the strength of the vortical structures inside the boundary layer only slightly as evidenced by the data in table \ref{tab:aerodynamics}. All values are reduced except for a minor increase of the Reynolds shear stress for PTE1. These observations suggest that the turbulence intensity is not distinctly increased - in contrast to other porous trailing edge designs~\cite{koh2018,teruna2021}. Hence, it is expected to cause only a negligible variation of the viscous drag. In fact, the larger recirculation region induced by the porous trailing edges (figure~\ref{fig:ptes_bl}~(b)) may result in a reduction of the overall wall-shear stress and skin-friction coefficient, which in turn would even lower the viscous drag.

\begin{table}[t]
\footnotesize
\centering
\caption{Median Reynolds shear stress $\overline{u'v'}$ and strength of the vortical structures $\bar{\lambda}$ inside the boundary layer downstream of the shock wave. The latter is determined based on the swirling strength, which uses the imaginary part of the eigenvalue of the velocity gradient tensor to reveal rotational flow structures like vortices~\cite{zhou1999}. The percentage difference to the values of the reference case is given in parentheses.}
\label{tab:aerodynamics}
\begin{tabular}{c||c|c}
 & $-\overline{u'v'}$ & $\bar{\lambda}$ \\
 \hline
REF & $194.5$ & $13787$ \\
PTE1 & $198.8 \quad (2.2~\% \uparrow)$ & $13331 \quad (3.3~\% \downarrow)$ \\
PTE2 & $193.4 \quad (0.6~\% \downarrow)$ & $13078 \quad (5.1~\% \downarrow)$ \\
\end{tabular}\\
\end{table}

Another significant drag source of aircraft operated at transonic speeds is wave drag. In this context, it arises due to the shock wave. When porous trailing edges are deployed, the averaged shock wave position is located slightly further downstream compared to the reference configuration (ref. table~\ref{tab:properties}). This can potentially result in a stronger shock wave which may induce an increased wave drag. However, since the downstream shifts are only $2.1\%$ (PTE1) and $1.0\%$ (PTE2), the associated aerodynamic penalty is expected to be marginal. \\[0.2cm]
Overall, we observe a reduced fluctuation of the studied quantities, e.g., shock wave motion and boundary layer breathing (ref. table~\ref{tab:properties}). This stabilizes the flow field and reduces the unsteadiness of the aerodynamic forces, which indicates a favorable effect on the global aerodynamic performance. Hence, the applied design choices might be able to counteract the aerodynamic penalties previously reported in the literature and avoid aerodynamic performance deficits. 

\subsection{Physical mechanism underlying the buffet mitigation effect of porous trailing edges} \label{sec:mechanism}
\begin{figure}[t!]
\includegraphics[width=1.\textwidth]{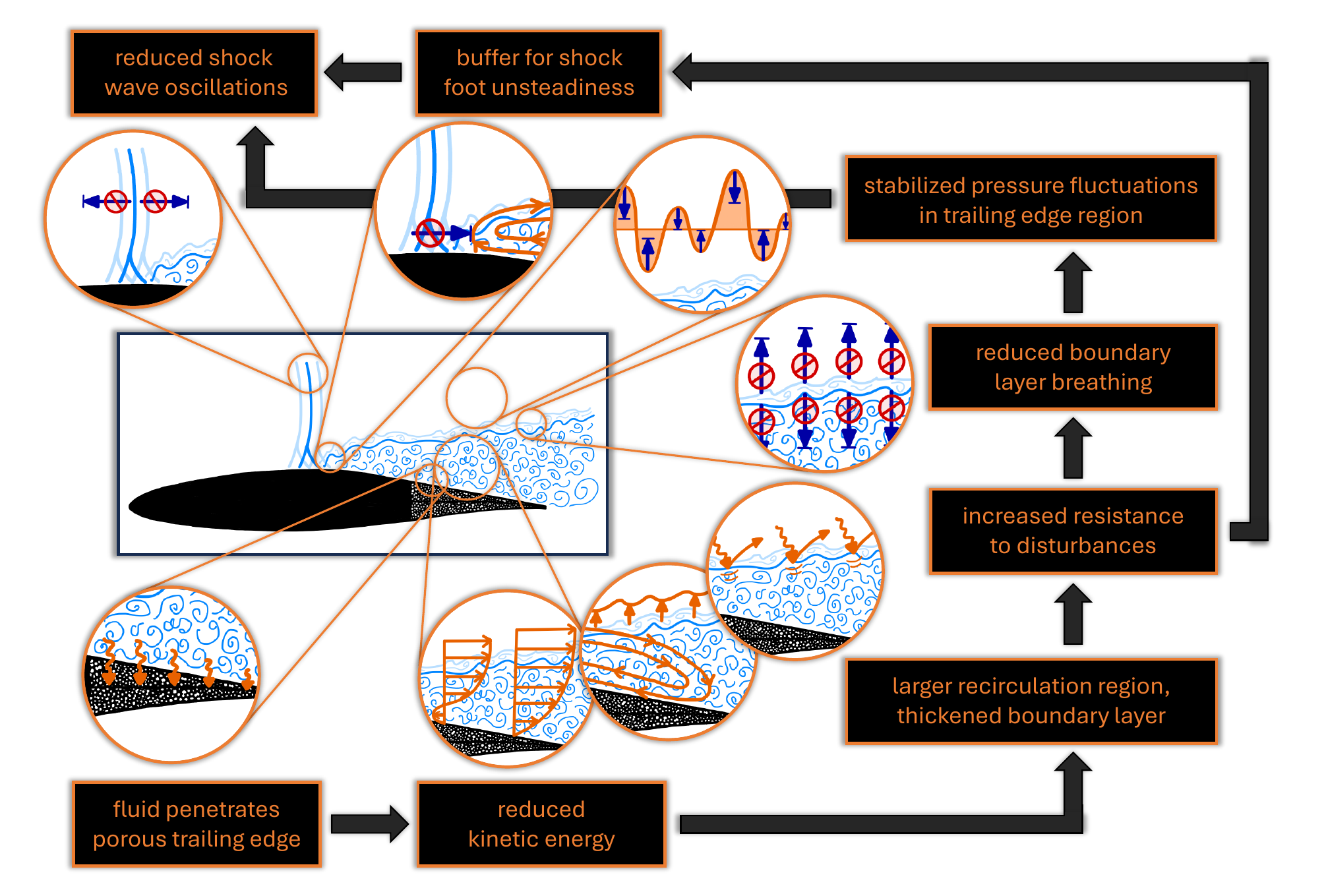}
  \caption{\textbf{Schematic diagram explaining how porous trailing edges interact with the flow field to mitigate the shock wave oscillations of transonic airfoil buffet.} In the presence of PTEs, fluid penetrates the porous material, which results in a reduction of the kinetic energy of the flow. Consequently, the recirculation region enlarges and the boundary layer thickness increases. Therefore, the boundary layer has an increased resistance to incoming disturbances, which provokes two different effects. First, the boundary layer breathing is reduced, which damps the pressure fluctuations in the trailing edge region. Second, the recirculation region acts as a buffer at the shock foot, i.e., it damps instabilities associated with the shock wave. Both effects cause an overall reduction of the shock wave oscillations. Consequently, the physical connection between the boundary layer downstream of the shock wave and the shock wave itself enables a mitigation of the shock wave movement via porous trailing edges, although the location of such trailing edge devices inhibits a direct interaction with the shock wave.\\\hspace{\textwidth} Please note that the physical processes sketched in this figure are simplified and exaggerated to illustrate the respective phenomena more clearly.}
  \label{fig:mechanism}
\end{figure}

The experimental studies demonstrate that both investigated porous trailing edge designs substantially attenuate the adverse shock wave oscillations associated with transonic airfoil buffet. In the following, we will explain how the PTEs interfere with the flow field to mitigate the shock wave movement since their location does not allow a direct interaction with the shock wave itself. The explanation is illustrated by the schematic diagram given in figure~\ref{fig:mechanism}. Essentially, the success of PTEs is rooted in the physical coupling of the shock wave oscillations and the boundary layer thickness variation, the latter being directly influenced by the PTEs.

When replacing a solid surface by a porous material, fluid penetrates the porous structure. This is known as the transpiration effect, which effectively reduces the kinetic energy of the fluid flow in the vicinity of the porous surface. In the present scenario, this effect can be observed based on the altered velocity distributions in the boundary layer (figure~\ref{fig:ptes_bl}~(b,c)). The streamwise velocity data additionally reveal an extended recirculation region which, in combination with the reduced kinetic energy close to the wall, results in a thicker boundary layer (figure~\ref{fig:ptes_bl}~(a)).

A thicker boundary layer possesses an increased resistance to perturbations. Incoming disturbances are more efficiently damped, which reduces the boundary layer breathing as depicted in figure~\ref{fig:ptes_bl}~(a) and table~\ref{tab:properties}. In turn, the attenuated upwards and downwards movement of the boundary layer stabilizes (i.e. reduces) the pressure fluctuations above the rear part of the airfoil. Since such pressure fluctuations directly affect the shock wave position, their damping results in a reduction of the shock wave movement. As a second contributor, the extended recirculation region behind the shock wave (figure~\ref{fig:ptes_bl}~(b)) stabilizes the shock wave by reducing the unsteadiness associated with the interaction region. It basically acts as a buffer absorbing perturbations that would otherwise provoke a shock wave movement. These connected phenomena explain why porous trailing edges are effective devices for transonic buffet attenuation although their location does not allow a direct interaction with the shock wave. This explanation is supported by studies from the literature which identify the shock foot instability and the connection between shock wave oscillation and boundary layer breathing as the driving mechanisms for transonic airfoil buffet, e.g., \cite{iwatani2023,paladini2019various,poplingher2019modal,fukushima2018wall,iovnovich2012reynolds}.  \\[0.2cm]
In addition to these observations, we noticed that the streamwise velocity distribution inside the boundary layer in the presence of PTEs is comparable to the distribution in a pre-buffet configuration (figure~\ref{fig:ptes_bl}~(d), $Ma \le 0.72$). However, the averaged shock position, the averaged boundary layer thickness, and the vertical velocity distribution inside the boundary layer differ from the pre-buffet conditions. Thus, the streamwise velocity distribution inside the boundary layer is an essential feature for transonic airfoil buffet. Consequently, its modification can attenuate the buffet phenomenon due to the direct coupling between the boundary layer properties and the shock wave movement as described above.\\[0.2cm]
Finally, the minor differences between the two porous trailing edge designs, PTE1 and PTE2, are addressed. The thickened boundary layer in the presence of PTEs effectively changes the surface contour imposing a favorable pressure gradient. This means that the pressure in the trailing edge region is reduced, which shifts the average shock position downstream as observed in figure~\ref{fig:ptes_shock} and table~\ref{tab:properties}. Since the boundary layer thickening of PTE2 is less intense, the shock position downstream shift is also less severe. Another effect of the thickened boundary layer is its increased resistance with respect to perturbations. The varying damping efficiency of both PTEs regarding the boundary layer breathing can be traced back to the different porous structures of the PTEs. Therefore, PTE2 reduces the fluctuation of the boundary layer thickness more efficiently compared to PTE1. 

\section{Discussion} \label{sec:discussion}
Porous trailing edges have attracted substantial interest in recent years since they lower acoustic aircraft emissions~\cite{teruna2020,sarradj2007,herr2014}. The present study reveals a second benefit of PTEs - their ability to attenuate self-sustained shock wave oscillations associated with the transonic airfoil buffet phenomenon. This result is important for the aviation sector since it demonstrates a promising strategy to extend the high-speed limit of the flight envelope and simultaneously reduce aircraft noise. Moreover, the results of this study provide new insight into the physical mechanism underlying the buffet phenomenon. This knowledge allows the design of efficient PTEs for prospective aircraft implementations that are particularly tailored to the expected flow conditions. \\
In summary, this experimental study shows that PTEs modify the velocity distribution within the separated boundary layer downstream of the shock wave. Since the altered streamwise velocity mimics the characteristics of pre-buffet flow conditions, its modification via the PTEs has an essential impact on the shock wave oscillations. On the one hand, the extended recirculation region damps instabilities associated with the shock-wave/boundary-layer interaction. On the other hand, the boundary layer thickening increases its resistance with respect to disturbances resulting in a less intense boundary layer breathing. This reduces pressure fluctuations in the trailing edge region with a direct impact on the shock wave oscillation due to the inherent connection between the shock wave position and the downstream pressure. In other words, a reduction of downstream pressure variations stabilizes the shock wave and damps transonic airfoil buffet. \\
The present investigation does not allow comprehensive conclusions about the maximum efficiency of PTEs in the context of buffet mitigation since only two PTE designs have been tested under fully developed buffet conditions. However, the manufacturing of PTEs via selective laser melting processes allows a flexible and fast design facilitating prospective parameter studies with different porous structures. In addition to analyzing the full potential of PTEs in the context of fully developed buffet, it is also interesting to assess how PTEs influence the buffet onset conditions, i.e., how far they can shift the occurrence of shock wave instabilities to higher Mach numbers and/or angles of attack. Moreover, the capabilities of the experimental facility allow a thorough analysis at higher air speeds to investigate the (potentially) upper limit of the new flight envelope. Complementary high-fidelity numerical simulations could provide additional aerodynamic quantities like lift and drag, which cannot be measured experimentally in the wind tunnel facility. These data can be used to validate our theoretical considerations regarding the aerodynamic performance in the presence of PTEs. In this context, we aim to incorporate the benefits of machine learning based data assimilation techniques leveraging the power of latent dynamical models \cite{lagemann2023learning} to extract physical mechanisms. We further intend to employ causal learning frameworks \cite{lagemann2023deep} that directly allow us to derive the causal relations underlying the flow system.

\section{Methods}\label{sec:Materials}
\subsection{Experimental setup}\label{sec:methods-experimental_setup}
The experimental measurements are conducted in the \textit{Trisonic Wind Tunnel} facility at the Institute of Aerodynamics of RWTH Aachen University, which operates at subsonic, transonic, and supersonic conditions. The present investigations are conducted within the transonic measurement section at Mach numbers in the range of $Ma = \{0.70; 0.76\}$. The main body of research focuses on measurements at $Ma = 0.76$ since these flow conditions exhibit fully developed buffet conditions. Only the reference configuration with an ordinary solid trailing edge is additionally investigated at lower Mach numbers. These data are compared with the modified flow topology induced by the porous trailing edges to assess if these devices transfer the flow field into a pre-buffet condition. \\
The experimental airfoil model is based on a supercritical DRA $2303$ profile, which was extensively investigated in former experimental and numerical studies~\cite{hartmann2012, hartmann2013interaction, hartmann2013coupled, feldhusen2018investigation, feldhusen2021analysis}. It has a relative thickness to chord ratio of $14\%$ with a total chord length of $c = 150$~mm. The leading edge part and the solid trailing edge are made of stainless steel while the porous trailing edges are made of titanium. A zigzag strip is glued onto the airfoil model at about $5\%$ chord to trigger a boundary layer transition. This is necessary to simulate a turbulent boundary layer flow similar to real flight conditions. \\
The airfoil model is installed in the measurement section at a fixed angle of attack of $\alpha = 3.5^\circ$ relative to the freestream velocity. The Reynolds number based on the freestream velocity and the chord length of the airfoil is $Re_c = 2.1\cdot 10^6$. Since the wind tunnel is operated at ambient flow conditions, Reynolds number and Mach number are linked via the fluid state and cannot be adjusted separately.\\
To exclude any influence of the wind tunnel walls on the flow field around the airfoil model, the measurement section has adjustable side walls. They are calibrated based on 26 dynamic pressure transducers distributed along the centerline of the upper and lower wall such that the wall curvature follows exactly the streamlines, which simulates an unbounded environment like in real flight conditions. \\
The wind tunnel itself is an intermittently working vacuum storage facility in which the fluid flow is generated by an extraordinarily high pressure difference. A massive reservoir tank downstream of the measurement section is evacuated, while the dehumidified air is stored in a big balloon upstream of the measurement section. A silica gel based drier ensures a relative humidity of the air below $4~\%$ at a total temperature of about $293$~K~\cite{hartmann2013interaction} such that no condensed water impedes the measurements. By opening a shutter, the pressure difference between evacuated tank and balloon creates an airflow from the balloon into the tank through the measurement section. Thereby, stable measurement conditions are present for about $2$ seconds in each measurement run. Due to limited storage capacity of the high-speed cameras, the duration of one measurement is limited to about $85$ buffet cycles.  

To simultaneously capture the shock wave motion and the fluid flow in the trailing edge region, synchronized Particle-Image Velocimetry (PIV) and Background-Oriented Schlieren (BOS) measurements are conducted. The BOS data are acquired with two \textit{Photron Fastcam SA5} cameras equipped with  \textit{180~mm Tamron f/8} lenses. A high-power LED illuminates a random dot pattern, which is recorded by the cameras at a frame rate of $8000$~Hz. Any density gradient changes in the flow field create an optical distortion of the dot pattern, which is captured in the BOS images. Since a shock wave constitutes a discontinuous density jump, the shock wave oscillation can be precisely studied via the BOS setup. \\
To capture the velocity field, simultaneous PIV measurements are conducted. Therefore, Di-Ethyl-Hexyl-Sebacat (DEHS) particles are added to the airflow directly within the balloon. The particles' properties are matched to the flow properties such that they perfectly follow the fluid flow. They are illuminated by a pulsed \textit{Darwin Duo 527-100-M} laser at $4000$~Hz and captured by two \textit{pco.dimax HS4} high-speed cameras equipped with \textit{100~mm Makro-Planar T* ZF.2} lenses. Due to limited optical access, the PIV cameras are setup at an angle relative to the laser light sheet (see figure~\ref{fig:setup}). Scheimpflug adapters are used to compensate this shift in perspective. 

The porous trailing edges are manufactured using selective laser melting processes, which is a 3D printing technique based on melting and recombining metallic powder using a high power laser. The two porous trailing edges used in the present study differ in the functional structure and permeability. The first trailing edge (PTE1) has a latticed structure with smaller pore diameters and lower permeability compared to the second trailing edge (PTE2). The second device consists of stacked gyroid cubes resulting in larger pore diameters and a higher permeability. Its composition makes it mechanically more stable, which is an important factor for prospective real-world applications. To prevent undesired flow perturbations at sparsely connected struts close to the trailing edges' surface, a perforated surface layer is added to both PTEs as shown in figure \ref{fig:aerodynamics}. Furthermore, a solid plate is included along the centerline of the trailing edge profile to inhibit a mass flux and thus, a pressure compensation, between the suction and the pressure side of the airfoil, which would degrade the aerodynamic performance.

\subsection{Data evaluation}
A state-of-the-art in-house code~\cite{marquardt2019, marquardt2020experimental} is used to evaluate the PIV data. The data consist of double images, i.e., two consecutive images acquired within a short time frame, which are used to estimate the particles' displacement in a statistical sense via cross-correlation. The known time interval between the two images is then used to transfer the horizontal and vertical displacement field into horizontal and vertical velocity information. The whole evaluation procedure includes sophisticated state-of-the-art processing tools to achieve the highest physical accuracy of the resulting velocity fields. That is, a multi-grid approach with steps $8-4-2$ is deployed, where the size of the interrogation windows is progressively reduced until the final window size $16 \times 16$~px$^2$ with an overlap of $75~\%$ is reached. A five-step predictor-corrector scheme incorporated into an iterative process allows for the highest accuracy in the displacement estimation. A Gaussian peak fit estimator enables sub-pixel accurate estimates, while a normalized median test reliably detects outliers. More details about the evaluation scheme are provided in \cite{marquardt2019,marquardt2020experimental,lagemann2023uncovering}.

The BOS data are evaluated with a deep optical flow network called RAFT-PIV~\cite{lagemann2021,lagemann2022}, which was recently proven to provide accurate density gradient estimates from BOS data~\cite{lagemann2022lxlaser}. In principle, each BOS image is compared to the initial reference image at zero flow to extract the changes in the density gradient due to the fluid flow. Compared to established evaluation routines, RAFT-PIV provides a flow state estimate for each pixel. This massively increases the accuracy and the spatial resolution of the final output compared to cross-correlation based methods. It is especially helpful in the present context to precisely extract the shock position. Since the data are not averaged across the finite size evaluation window like in classical approaches, the shock localization is significantly more precise. Moreover, we further enhance the shock detection by applying the 2D Noise-Assisted Multivariate Empirical Mode Decomposition (2D NA-MEMD)~\cite{maeteling2022} to the density gradient fields provided by RAFT-PIV. This data-driven method extracts scale-based modal representations from the input data in a spatio-temporal framework based on the scales inherent to the data. The sharp edges of the shock wave distinctly appear in the first mode, which allows an accurate extraction of the shock wave position for each individual time instant. A temporal coherence is achieved by simultaneously decomposing consecutive snapshots of the flow field. This temporal consistency is further enhanced by deploying the concept of noise assistance, where we use two additional channels containing random Gaussian noise with a standard deviation of $2$~\% of the standard deviation of the multivariate data in the decomposition. The reader is referred to \cite{maeteling2022,maeteling2023,lagemann2023impact,lagemann2022lxlaser} for further details on the 2D NA-MEMD and its application to fluid flow data.

\bibliographystyle{ieeetr}

\end{document}